\documentclass{pasj00}
\begin{document}
\SetRunningHead{Gopalswamy \& Yashiro}{Obscuration by an Eruptive Prominence}
%\Received{}%{yyyy/mm/dd}
%\Accepted{}%{yyyy/mm/dd}
%\Published{}%{yyyy/mm/dd}

\title{Obscuration of Flare Emission by an Eruptive Prominence}

%%% begin:list of authors
% Do NOT capitalize all letters in "textsc".

%%% end:list of authors

%%% Please use the following style in case that sorting by
%%% affiliation is impossible.
%
\author{%
  Nat \textsc{Gopalswamy}\altaffilmark{1}
  and
  Seiji \textsc{Yashiro}\altaffilmark{1,2}}
\altaffiltext{1}{Code 671, NASA Goddard Space Flight Center, Greenbelt, MD 20771, USA}

\altaffiltext{2}{Department of Physics, The Catholic University of America, Washington DC 20064 USA}
\email{nat.gopalswamy@nasa.gov; seiji.yashiro@nasa.gov}

%% `\KeyWords{}' always has to be placed before `\maketitle'.
 %Do NOT move this preamble from here!

\maketitle

\begin{abstract}
  We report on the eclipsing of microwave flare emission by an
  eruptive prominence from a neighboring region as observed by the
  Nobeyama Radioheliograph at 17 GHz. The obscuration of the flare
  emission appears as a dimming feature in the microwave flare light
  curve. We use the dimming feature to derive the temperature of the
  prominence and the distribution of heating along the length of the
  filament. We find that the prominence is heated to a temperature
  above the quiet Sun temperature at 17 GHz. The duration of the
  dimming is the time taken by the eruptive prominence in passing over
  the flaring region. We also find evidence for the obscuration in EUV
  images obtained by the Solar and Heliospheric Observatory (SOHO)
  mission.
\end{abstract}

\section{Introduction}

One of the greatest contributions of the Nobeyama radioheliograph
(NoRH, Nakajima et~al. 1994) to the study of solar eruptive phenomena
is the information on the eruption of filaments and prominences, which
are near-surface manifestations of coronal mass ejections (CMEs).
Investigations involving Nobeyama prominence eruptions and white-light
CMEs have clarified a number of issues and enhanced our understanding
of solar eruptions (see e.g., Hanaoka et~al. 1994; Gopalswamy and
Hanaoka 1998; Gopalswamy et~al. 1998; 1996; Hanaoka and Shinkawa,
1999; Hori and Culhane, 2002; Gopalswamy and Thompson, 2000;
Gopalswamy et~al. 2003a; Kundu et~al. 2004; Shimojo et~al. 2006;
Gopalswamy et~al. 2012).  By tracking the locations of prominence
eruptions as a function of time, it was found that the polarity
reversal at solar poles coincided with the times of cessation of high
latitude (poleward of 60$^\circ$) eruptions (Gopalswamy et~al. 2003b;
2012).  The spatial relationship between prominence eruptions and the
associated CMEs has also been useful in refining our understanding of
CME deflection by coronal holes, especially with respect to the
evolution of the poloidal field of the Sun (Gopalswamy et~al. 2000;
Gopalswamy and Thompson, 2000). In particular, it was possible to show
that the position angle of prominence eruptions is systematically
offset poleward from the CME position angle during the rise phase of
solar cycles (Gopalswamy et~al. 2003a; 2012). This offset becomes
random in the maximum phase because the polar coronal holes are
present only up to the beginning of the maximum phase.

In this paper, we report on another aspect of prominence eruptions:
obscuration of neighboring bright flare structure by an eruptive
prominence. The observation provides quantitative information on the
heating of the prominence during its eruption.  We also show that a
larger volume of higher optical depth plasma surrounds the eruptive
filament.

\begin{figure*}
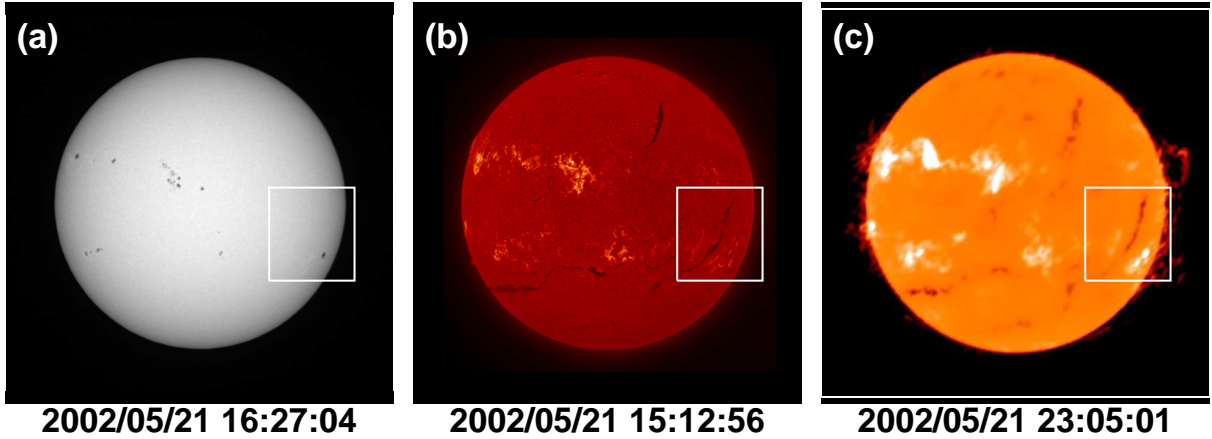

  \begin{center}
    \FigureFile(160mm,10mm){fig1.eps}
    %%% \FigureFile(width,height){filename}
  \end{center}
  \caption{White light 16:27:05 UT (a) and H-alpha at 15:12:56 UT on
  May 21 (b) images from the Big Bear Solar Observatory, and 17 GHz
  microwave image from the Nobeyama radioheliograph at 23:05 UT
  showing the region of interest. The Big Bear H-alpha and white light
  images show the large sunspot in the negative polarity region of AR
  9948.}\label{fig:context}
\end{figure*}

\begin{figure*}
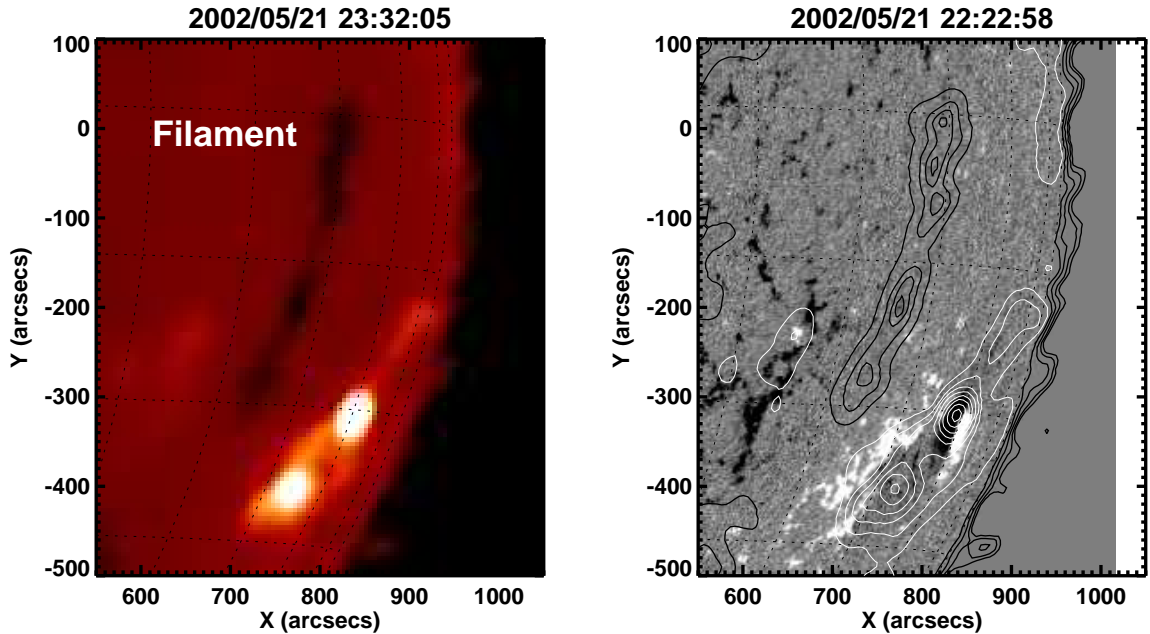

  \begin{center}
    \FigureFile(160mm,10mm){fig2.eps}
    %%% \FigureFile(width,height){filename}
  \end{center}
  \caption{Overlay of microwave contours on the SOHO/MDI magnetogram
  showing Flare1 and the filament. The filament eruption was
  associated with a spotless flare and a fast CME. The contour levels
  are 6000, 7000, 8500, 10000, 12000, 15000, 20000, 30000, 50000,
  80000, 130000, 180000~K. In the case of the black (white) contours
  the brightness temperature decreases (increases) towards the
  peak. The sunspot source has the highest Tb (180,000~K). The
  southern flare source has a Tb of 130,000~K. The lowest temperature
  in the filament is 6000~K.}\label{fig:contour}
\end{figure*}

\section{Observational Overview}

We are concerned about two flares that occurred in the southwest
quadrant of the Sun during 2002 May 21-22. The first one was a C9.7
flares from active region NOAA 9948 (S25W64) followed by a C5.0 flare
from a neighboring filament region (S30W34). A long quiescent
filament, which erupted in association with the C5.0 flare, obscured
the preceding flare emission as it passed over it. A partial halo CME
with a speed of 1246 km/s was associated with the C9.7 flare, while a
full halo CME with a speed of 1557 km/s was associated with the C5.0
flare. The second CME produced an interplanetary type II radio burst
and a large solar energetic particle event.  Figure \ref{fig:context}
shows the region of interest consisting of AR 9948 with a large
sunspot (Fig. 1a) and a neighboring filament observed in H-alpha
(Fig. 1b) and in microwaves by NoRH (Fig. 1c). The filament extended
slightly beyond the lower left corner of the box shown in Fig. 1. The
filament was located along the neutral line of an extended plage
region to the east of AR9948. The sunspot in AR 9948 appears bright in
microwaves (Fig.1c) because of the thermal gyroresonance emission due to
the high magnetic field above the sunspot located in the negative
polarity region.  The C9.7 flare, which started, peaked, and ended at
23:14, 00:30, and 01:28 UT, respectively involved the sunspot also.

\begin{figure*}
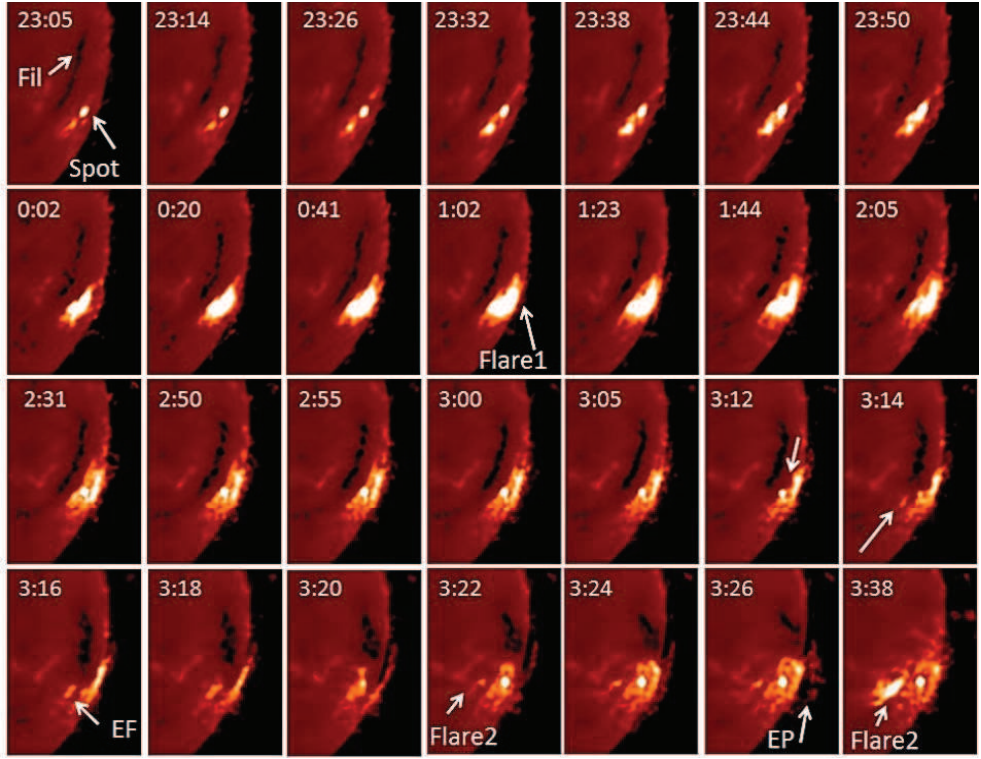

  \begin{center}
    \FigureFile(130mm,10mm){fig3.eps}
    %%% \FigureFile(width,height){filename}
  \end{center}
  \caption{Evolution of the two flares in question (Flare1 and Flare2)
  along with the eruptive filament from 23:05 UT on 2002 May 21 to
  03:38 UT on May 22. The filament (Fil) and the sunspot in the
  pre-eruption state are marked in the 23:05 UT (May 21) image. The
  post-eruption arcade of Flare1 is marked in the 01:02 UT (May 22)
  image. The first instance of significant obscuration by the filament
  is at the location indicated by the arrow in the 03:12 UT image. In
  the images at 01:14 and 03:16 UT, the eruptive filament (EF) divides
  the flare emission into two parts. The flare emission starts
  recovering from 03:20 UT onwards and the eruptive filament becomes
  an eruptive prominence (EP) at 03:26 UT.  The last image at 03:38 UT
  shows the post-eruption arcade corresponding to Flare2. The compact
  source is due to the Sunspot. In microwaves, the sunspots appear
  bright because of the thermal gyro-resonance
  emission.}\label{fig:evolution}
\end{figure*}

Figure \ref{fig:contour} provides a detailed view of the flaring
regions and their magnetic field configuration.  The NoRH snapshot at
23:32:05 UT on May 21 was taken about 18 min after the flare
onset. The image shows the compact sunspot source at (850,-320) and a
new source in the south (770,-400). The brightness temperatures
(Tb) of the northern (sunspot) and southern sources were
$1.8\times10^5$~K and $1.3\times10^5$~K, respectively. The sunspot
source attained a highest Tb of $4.9\times10^5$~K during its peak at
00:17:05 UT on May 22, about 13 min before the flare peak in soft
X-rays.  The thermal component of the flare was in the form of a
post-eruption arcade, which had an average Tb in the range
$2-3\times10^4$~K (more details in the next section). The thermal
emission encompassed the two compact sources and occupied a much
larger volume. The filament and the compact sources are shown overlaid
as contours on the nearest SOHO/MDI magnetogram taken at 22:22:58 UT
on May 21 and rotated to the time of the NoRH image. The two compact
sources are on opposite polarity regions. The filament in question
divides the opposite polarity regions of the large plage region to the
east of the sunspot.

The lowest brightness temperature was $\sim6000$~K in the darkest
parts of the filament shown in Fig. 2. The quiet-Sun Tb is $\sim
10^4$~K, corresponding to the level between the highest dark contour
($10^4$~K) and the lowest white contour ($1.2\times10^4$~K). The
second flare started near the southern end of the filament, where the
filament had a higher Tb, close to the quiet Sun level.  The filament
started lifting off slowly around 03:12 UT on May 22 and the soft
X-ray flare started around 3:18 UT. The first flare was in the decay
phase at this time, but the post eruption arcade was clearly observed
in microwaves.  The northern leg of the filament remained anchored
throughout the eruption.  Between 03:12 and 03:24 UT, the eruptive
filament passed over the arcade of the first flare causing the dimming
we are interested in.

\begin{figure*}
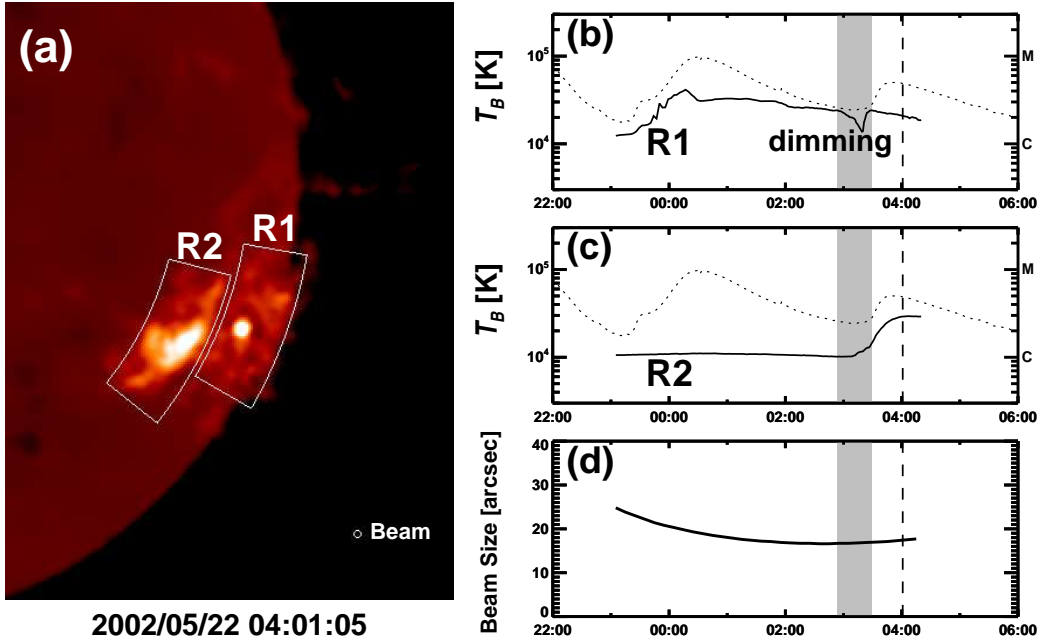

  \begin{center}
    \FigureFile(140mm,10mm){fig4.eps}
    %%% \FigureFile(width,height){filename}
  \end{center}
  \caption{(a) Regions R1 and R2 enclosing microwave flare emission
  from Flare1 and Flare2, respectively. In the image shown, R1 has the
  decayed post-eruption arcade from Flare1 and the bright sunspot
  source. In R2, the bright feature is Flare2. (b,c) Evolution of the
  flare Tb from 22:00 UT on 2002 May 21 to 06:00 UT on May 22 in R1
  and R2, respectively. The duration of the dimming in (b) is
  indicated by the gray bar. The GOES light curves of the two flares
  are shown by the dotted line (the flare intensity level is indicated
  on the right-side Y-axis). The soft X-ray emission before 23:00 UT
  is from an unrelated flare in the northeast quadrant.  The GOES soft
  X-ray observation has no spatial information, so the flaring regions
  cannot be isolated as is done for the microwave emission. The area
  of the boxes is 1843 pixels ($2.3\times10^10$~km$^2$). The pixel
  area is $4.91\times4.91$ ($5.2\times10^5$~km$^2$). The vertical
  dashed line in (b) and (c) marks the time of the NoRH image in
  (a). In (d), the evolution of the NoRH beam size is shown. The beam
  size had a constant value of $\sim16''.8$ during the time of the
  obscuration. }\label{fig:regions}
\end{figure*}

\section{Detailed Evolution of the Two Flaring Regions}

Figure \ref{fig:evolution} shows a series of NoRH images between 23:05
UT on May 21 to 03:38 UT on May 23 illustrating the evolution of the
two flares and the eruptive filament eclipsing the microwave emission
from the first flare.  In the pre-flare state, all we see is the
bright sunspot and some plage brightening to the south of the spot
(23:08 UT). When the flare started, new compact footpoint sources
appeared in the south and at the southern edge of the sunspot (22:32
UT). In the next few minutes, the flare emission appeared in the
region between the two compact sources and to the north of the
sunspot. Between 00:02 and 01:44 UT, the flare arcade grew in size and
the sunspot emission was no longer bright enough to be seen
distinctly. When the flare arcade started fading by 02:31 UT, the
sunspot emission can be seen again as a compact source. Around 03:05
UT, the southern end of the filament started lifting and by 03:12 UT
the eruptive filament (EF) appeared orienting north-south due to the
eruption. Even though the EF was still to the east of the flaring
region, the flare emission was already dimmed due to the material
surrounding the darkest part of the filament. The dimmed region is
pointed by an arrow in the 03:12 UT image. In the next two minutes,
the sunspot emission was completely eclipsed by the EF and the flare
emission appears divided by the EF. The eastern edge of the flare
emission started reappearing after the passage of the EF over it as
pointed by an arrow in the 03:14 UT image. In the 03:16 UT image, the
EF roughly bisected the flare emission into eastern and western
halves. The southern part of the EF was clearly thinner, but much
wider over the sunspot.  By this time, the EF was in the north-south
direction and the darkest part of EF had shortened
significantly. Furthermore, the EF appeared much wider than the
initial thickness probably because the viewing angle changed due to
the moving filament. The maximum obscuration of the flare emission
occurred around 03:18 UT. By 03:20 UT, the sunspot emission started
emerging from the EF eclipse and by 03:22 UT, most of the flare
emission reappeared except for a small section closer to the limb. By
03:24 UT, the entire flare emission reappeared and the EF became
eruptive prominence (EP) as it appeared above the limb. NoRH was able
to track the EP to much larger distances into the corona and became
the core of the fast CME.  The second flare (Flare2) associated with
the EF became visible as a thin ribbon in the 03:22 UT NoRH image and
became the brightest structure by 03:38 UT.

\subsection{Time profile of the dimming}

Figure \ref{fig:regions} shows the variation of the brightness
temperature in two boxes that enclose the two flaring regions (R1, R2)
corresponding to the two flares.  The peak Tb in R1 was
$4.1\times10^4$~K at 00:17 UT, about 13 min before the soft X-ray
flare peak. After this time, the flare emission was mostly from the
post-eruption arcade and remained roughly constant at an average Tb of
$2.4\times10^4$~K until about 02:53 UT, when the dimming started due
to the passage of the eruptive filament over R1. The dimming lasted
until about 03:29 UT, when the filament had crossed the flare area and
moved above the limb. Thus the dimming lasts for $\sim36$~min. At the
time of the deepest dimming, the Tb was $1.4\times10^4$~K, which is
more than the quiet-Sun Tb. When the dimming ended, the Tb in R1
returned back to the pre-eclipse level ($2.4\times10^4$~K). This also
confirms that the eruptive filament was just obscuring the
quasi-steady flare emission. Note that the Tb reduction in Fig. 4 is
averaged over the area of R1. The flare emission in R2 starts right
after the dimming ended in R1 along with the C5.0 flare in GOES soft
X-rays.

\begin{figure*}
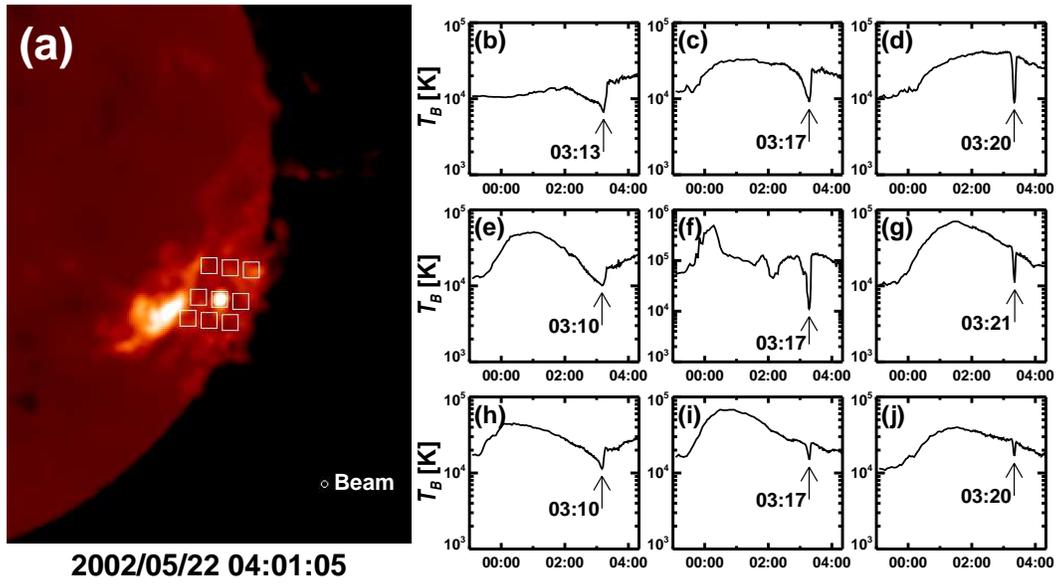

  \begin{center}
    \FigureFile(140mm,10mm){fig5.eps}
    %%% \FigureFile(width,height){filename}
  \end{center}
  \caption{Details of the obscuration at various locations in the
flaring region. (a) A NoRH image at 04:01 05 UT with several square
boxes overlaid. Each box has a side of 64 pixels ($2.26\times10^5$
km). The NoRH beam around the time of this image is also shown.  Time
evolution of the brightness temperatures in the nine boxes are shown
in the right: (b,c,d) $-$ top row, (e,f,g) $-$ middle row, and
(h,i,j) $-$ bottom row. The dimming of the microwave flare emission
is marked by arrows including the times when the dimming was the
deepest. The Y-axis scale is the same for all plots except for the one
in (f) to accommodate the high Tb in the vicinity of the sunspot. The
lower rows are shifted to the east to roughly align with the filament
orientation in the beginning.  The Tb increase in boxes (b), (e), and
(h) after the dimming is due to the extension of Flare2 emission into
these boxes.  Table 1. Times of deepest dimming and the corresponding
Tb in the boxes [b]-[j] in Fig. 5.}\label{fig:boxes}
\end{figure*}

\begin{table*}[bth]
  \caption{Times of deepest dimming and the corresponding Tb in the boxes [b]-[j] in Fig. 5}\label{tab:first}
  \begin{center}
    \begin{tabular}{llllll}
      \hline
Time & Tb Min ($10^4$~K) & Time & Tb Min ($10^4$~K) & Time & Tb Min ($10^4$~K)\\
\hline
03:13:05 & 0.7 [b] & 03:17:05 & 0.9 [c] & 03:20:05 & 0.9 [d] \\
03:10:05 & 1.0 [e] & 03:17:05 & 1.1 [f] & 03:21:05 & 1.1 [g] \\
03:10:05 & 1.1 [h] & 03:17:05 & 1.5 [i] & 03:20:05 & 1.7 [j] \\
\hline

    \end{tabular}
  \end{center}
\end{table*}

\subsection{Spatial Variation of the Flare Obscuration}

\begin{figure*}
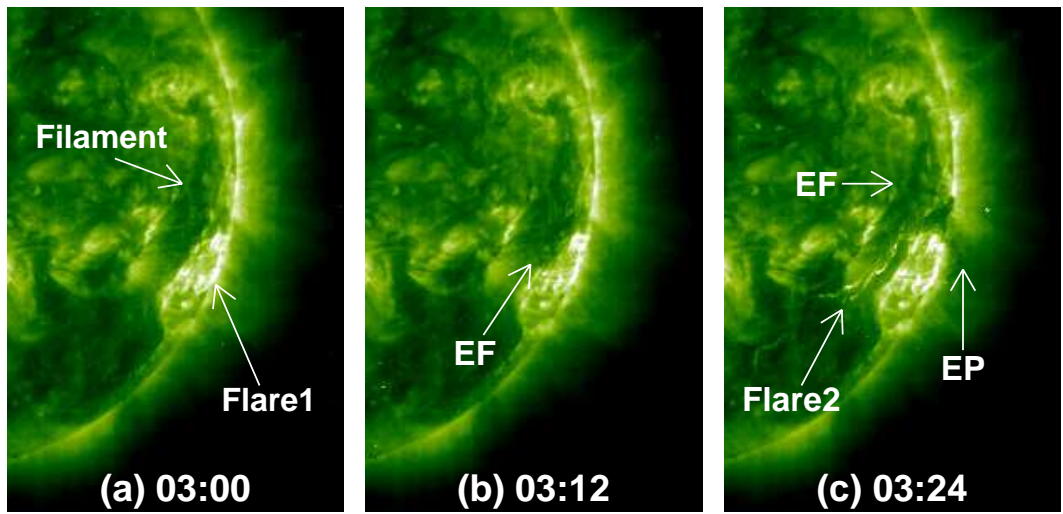

  \begin{center}
    \FigureFile(140mm,10mm){fig6.eps}
    %%% \FigureFile(width,height){filename}
  \end{center}
  \caption{SOHO/EIT images showing the obscuration of the flare
  emission (Flare1) by the eruptive filament (EF) from the location of
  the second flare (Flare2). The arrow in the middle panel shows the
  location where the EF obscured the flare emission.  In the right
  panel, the southern part of the EF moved above the limb and became
  an eruptive prominence (EP), while the northern leg is still on the
  disk (EF), similar to what was observed in
  microwaves.}\label{fig:euv}
\end{figure*}

In order to examine the spatial variation of the flare intensity due
to the eclipsing by the eruptive filament, we created 64-pixel boxes
within R1 as shown in Figure \ref{fig:boxes}, and tracked the Tb
variation within each of these boxes. Table 1 lists the times of the
deepest dimming in each box and the corresponding minimum Tb
values. The times of the deepest minimum is roughly the same for each
column and seems progressively later from left to right, tracking the
motion of the filament over the flaring region: 03:10 UT in the
eastern column (b,e,h), 03:17 UT in the middle column (c,f,i), and
03:20 UT in the western column (d,g,j).  The start time of the dimming
is not clear in the eastern column because it is located at the edge
of the flare arcade. The duration of the dimming ranged from 80 min to
9 min depending on the location. The dimming generally lasted longer
in the northern part where the filament and the surrounding structure
were larger. The duration of the dimming in the sunspot region was 30
min, similar to the average value in Figure \ref{fig:regions}. The
depth of the dimming essentially depended on the Tb of the obscuring
section of the filament. In the top row of boxes, Tb dropped to
$0.7\times10^4$~K to $0.9\times10^4$~K, remaining below the quiet Sun
level. This is also clear from the images in Fig. 3, which shows that
the EF remained dark in the north. In the bottom row of boxes, the Tb
dropped to $1.1\times10^4$~K to $1.7\times10^4$~K, which is
significantly higher than the quiet-Sun values. In the middle row, the
Tb dropped to the quiet Sun value or slightly higher
($1.0\times10^4$~K to $1.1\times10^4$~K). Since the EF completely
obscured the flare emission, the observed Tb is the same as the
filament Tb. When the filament material is optically thick, the
observed Tb is also the kinetic temperature of the filament (see
section 5 for a discussion on the optical thickness of the
filament). In other words, the filament was heated during the eruption
such that the temperature more than doubled in the southernmost part
(where the filament lifted off and Flare2 followed). In the middle
part, the filament temperature increased by about 50\% and in the
northern part, the filament temperature changed only marginally. There
is also a clear temperature gradient along the axis of the filament,
with the highest temperature near the eruption location and the lowest
temperature in the northern leg. This is consistent with the fact that
the northern leg of the filament remained dark until the filament
moved above the limb (see figure \ref{fig:evolution}).

It must be noted that the filament width is larger than the NoRH beam,
so the leakage of the bright flare into the beam is negligible. The
width of the filament at the thinnest part is $\sim62''.8$. The NoRH
beam size (full width at half maximum $-$ FWHM) is $\sim16''.8$
(see Figs. 4 and 5). Thus the leakage from the bright flare emission
into the Tb of the filament is expected to be very small, especially
to at the central part of the filament. The filament Tb at the
thinnest part is $\sim1.3\times10^4$~K.  On the east and west edges of
the filament, the flare emission had peak values of $2.3\times10^4$~K
and $2.4\times10^4$~K, respectively. The deepest part of the filament
is at a distance of $\sim32''.8$ and $\sim30''.0$ from the east
and west edges, respectively. By superposing the NoRH beam at these
edges (assuming Gaussian), we can see that the leakage from the bright
flare emission is negligible. The FWHM of $16''.8$ implies that the sigma
of the beam is $\sim7''.1$. Thus, the deepest part of the filament
is 4.6 sigmas away from the east edge, so the leakage component is
$\sim$0.5~K.  Similarly, the leakage contribution from the west edge
is $\sim3.2$~K. These values are much below the noise level on the maps.  The
finite size of the beam makes the size of the filament slightly
underestimated at the edges.

\begin{figure*}
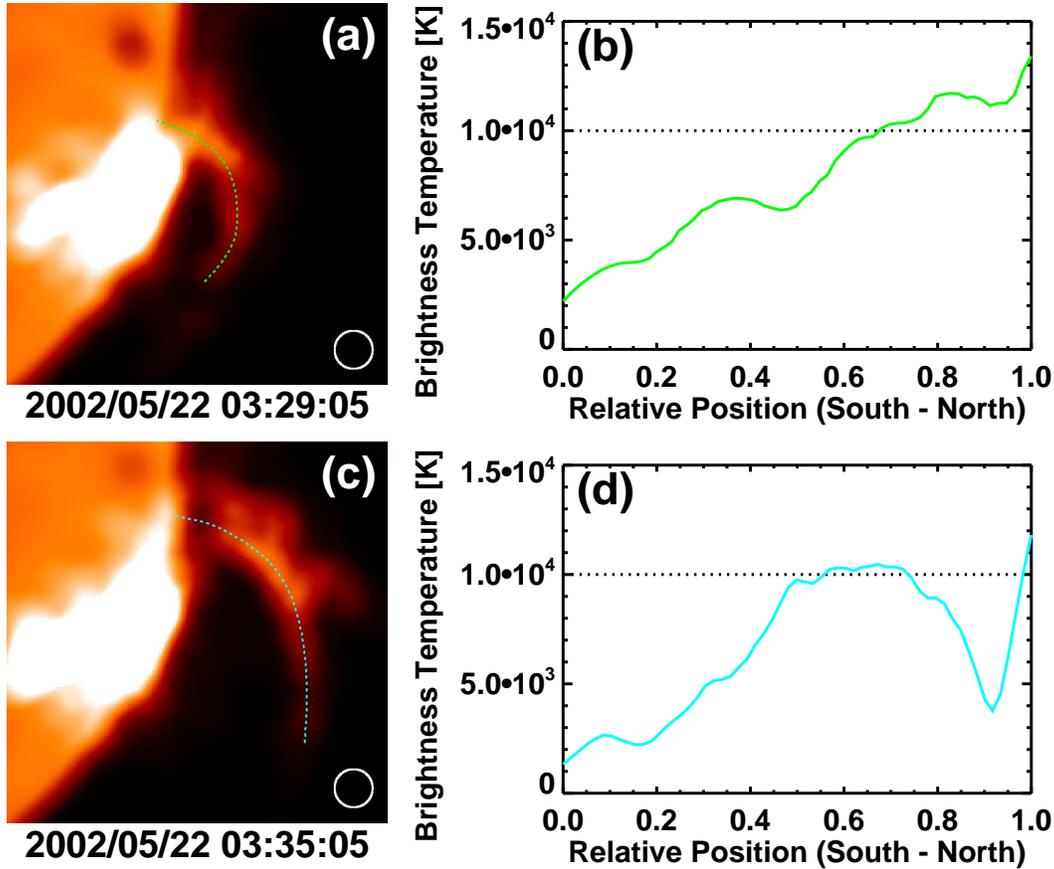

  \begin{center}
    \FigureFile(140mm,10mm){fig7.eps}
    %%% \FigureFile(width,height){filename}
  \end{center}
  \caption{Images of the eruptive prominence at two instances
  [03:29:05 UT (a) and 03:35:05 UT (c)] with the
  corresponding brightness temperature variations along the length of
  the prominence plotted in (b) and (d) from the southern end
  (position = 0) to the northern end (position = 1). The size of the
  circle (radius = 8 pixels) that was used in smoothing the images are
  shown in (a) and (c).  Note that some sections of the prominence
  have brightness temperatures exceeding $10^4$~K. }\label{fig:prom}
\end{figure*}

\section{Filament Obscuration in EUV}

In this section we present an independent confirmation of the dimming
of flare emission in EUV as observed in the images obtained by the
Extreme-ultraviolet Imaging Telescope (EIT) on board SOHO. The EIT
images were obtained with a low cadence (1 image every 12
minutes). Fortunately, there was one image at 03:12 UT, at which time
the dimming clearly was underway.  Figure \ref{fig:euv} shows three
EIT images taken at 03:00 UT, 03:12 UT, and 03:24 UT, roughly
corresponding to the pre-dimming, dimming, and post-dimming phases,
respectively. The partial obscuration of the EUV arcade emission is
clearly seen in the 03:12 UT image and the appearance is very similar
to that in the microwave image taken at the same time (see figure
\ref{fig:evolution}). In the image at 03:24 UT, the filament had
already passed over the flare and appeared above the limb. Note that
this dimming is due to obscuration by an eruptive filament from a
neighboring region and not the usual dimming one observes during the
eruption process, due to the evacuation of material on either side of
the neutral line.  Many eruptive events imaged by the Solar Dynamics
Observatory also show the obscuration-type dimming.

\section{Discussion}

In section 3.2 we had assumed that the filament was optically thick.
Here we show that the heated filament was optically thick when it was
obscuring the flare emission. The radiative transfer equation for the
observed brightness temperature can be written as (Dulk, 1985):
\begin{equation}
  Tb=\int_0^\tau T_F e^{-t} dt + T_a e^{-\tau}.
\end{equation}
Here $T_F$ and $T_a$ are the physical temperatures of the filament
and the flare arcade, respectively; $\tau$ is the optical depth of the
filament. For temperatures $T < 2\times10^5$~K, the absorption
coefficient is given by (Webb and Kundu, 1978),
\begin{equation}
  \kappa=0.014gn^2f^{-2} T^{-3/2}
\end{equation}
where $f$ is the observing frequency (17 GHz), $n$ is the plasma
density, and the Gaunt factor $g$ is given by,
\begin{equation}
g =18.2+\ln T^{3/2}-\ln f.
\end{equation}
For $T$ = $T_F$ = $1.3\times10^4$~K and $f$ = 17~GHz, we get $g$ = 8.8 and
\begin{equation}
\kappa = 2.9\times10^{-28} n^2.
\end{equation}
Assuming that the line of sight depth $L$ of the filament is similar
to the observed width ($62''.8$ or $4.5\times10^4$~km), we get the
optical depth $\tau$ = $\kappa$$L$ = $1.3\times10^{-18}n^2$. The
filament is optically thick if $n > 8.8\times10^8~$cm$^{-3}$.  This
inequality is readily satisfied because the density of the
pre-eruptive filament is typically $10^{11}$~cm$^{-3}$ and may not be
too different in the early phase we are investigating. Irimajiri et
al. (1995) found a density of a few times $10^{10}$~cm$^{-3}$ in a few
eruptive prominences observed by the 45~m telescope in Nobeyama. We
also have independent confirmation of the filament density from white
light observations made by the Solar and Heliospheric Observatory
(SOHO) mission's Large Angle and Spectrometric coronagraph. A
calibrated LASCO image at 04:50 UT was used for this purpose. The
eruptive filament became the CME core and was located at a distance of
2.7~Rs from the Sun center. The column density of the prominence
material was $1.3\times10^{18}$~cm$^{-2}$.  The filament had a width
of $8.6\times10^4$~km at the brightest part.  Assuming the line of
sight depth to be the same as the observed width, we obtained an
electron density of $1.5\times10^8$~cm$^{-3}$.  The filament might
have expanded and lost mass due to draining before arriving at 2.7~Rs
suggesting a reduction in density compared to near-Sun
values. Previous estimates have shown that the density decreases by
three orders of magnitude from near the Sun to the coronagraphic FOV
(Illing and Athay, 1986).  Thus the density of the filament at the
time of obscuration is expected to be $> 10^{10}$~cm$^{-3}$, so we
infer that the optical depth $>130$.  This result is also consistent
with the densities (2 - 5 $\times10^{10}$~cm$^{-3}$ derived by
Irimajiri et~al. (1995).  The observed Tb can be explained as a result
of heating of the whole filament or the filament with a cool core at
$\sim8000$~K with an optically thin heated sheath at a temperature of
$\sim10^5$~K (Hanaoka and Shinkawa, 1999).  The sheath needs to be
optically thin, contributing a few thousand K to the observed Tb. Note
that the second term in eq. (1) drops out due to the large $\tau$.

The heating of the eruptive filament is also evident when the filament
crossed the limb and became a bright eruptive prominence.  Although
some sections of the filament might have expanded and became optically
thin, the apex part of the prominence probably remained optically
thick. We infer this from the observed Tb.  Figure \ref{fig:prom}
shows the Tb variation along two arcs above the limb crossed by the
prominence. The arcs are at heliocentric distances of 1.2 and 1.4~Rs
in the sky plane. It is clear that some sections of the eruptive
prominence had Tb exceeding $10^4$~K. In order to avoid artifacts in
the observed prominence structure (see e.g., Koshiishi 2003), we
smoothed the image over circular areas with radius = 8 pixels ($39''$
or $2.8\times10^4$~km).  Thus the Tb plotted is an underestimate, but
confirms that there is definite heating. Let us take the peak value at
$1.2\times10^4$~K. The Tb is similar to that of the filament at the
time of obscuration.  An optically thick filament at 8000~K with an
optically thin contribution of $\sim4000$~K from the heated sheath is
consistent with the observations.  From equations (2) and (3), we get
$g$ = 11.9 and $\kappa$ = $1.8\times10^{-29}n^2$ for $T$ =
$10^5$~K. For an average sheath density $n$ = $10^9$~cm$^{-3}$, a
sheath thickness of $\sim2.2\times10^4$~km is needed to account for
the optically thin contribution ($\sim4000$~K) from the sheath.

\section{Summary}

The primary result of this paper is that an eruptive filament eclipsed
the flare emission for tens of minutes during its transit above the
post-eruption arcade in a neighboring active region. The duration of
the dimming roughly corresponds to the transit time of the eruptive
filament over the neighboring flare. The present observation provides
strong evidence that the eruptive filament was heated to temperatures
well above that of the quiescent filament. Furthermore, the filament
seems to be heated more near the eruption site and progressively less
away from the eruption site to almost no heating near the distant leg
of the filament. Some sections of the filament remained above 10$^4$~K
when the eruptive filament went above the limb as an eruptive
prominence. The increased brightness temperature can be explained by a
combination of the optically thick emission from the cool filament
core and an optically thin emission from the heated filament-corona
transition region at a temperature of $\sim10^4$~K. The dimming
observed is not the one produced as transient coronal holes on either
side of the neutral line, but a simple obscuration. The eruptive
filament overlies the hot flare plasma and hence completely eclipsed
by the filament. Note that the higher brightness temperature of the
flare plasma background made the heated filament visible at
microwaves.

%%%%%%%%%%%%%%%%%%%%%%%%%%%%%%%%%%%%%%%

\bigskip

Acknowledgement. The authors thank the Local Organizing Committee of
the ``Solar Physics with Radio Observations'' meeting for travel
support. The authors benefited greatly from the open data policy of
the Nobeyama Solar Observatory and NASA. SOHO is a project of
international cooperation between ESA and NASA.

%%%
% See the manual for the detail.
%%%

\end{document}